# Chatbot apologies: Beyond bullshit

P.D. Magnus, Alessandra Buccella, and Jason D'Cruz[1]

**ABSTRACT:** Apologies serve essential functions for moral agents such as expressing remorse, taking responsibility, and repairing trust. LLM-based chatbots routinely produce output that has the linguistic form of an apology. However, they do this simply because they are echoing the kinds of things that humans say. Moreover, there are reasons to think that chatbots are not the kind of linguistic or moral agents capable of apology. To put the point bluntly: Chatbot apologies are bullshit. This paper explores this concern and develops it beyond the epithet, drawing on the nature of morally serious apologies, the linguistic agency required to perform them, and the moral agency required for them to matter. We conclude by considering some consequences for how chatbots should be designed and how we ought to think about them.

---

Especially since the release of ChatGPT in late 2022, there has been a furor about chatbots powered by Large Language Models (LLMs).[2] Much of the concern has been directed at the problem of hallucination or confabulation, the tendency of chatbots to produce outputs which look like assertions but which have no connection to the truth. It is common to suggest that the output of chatbots is *bullshit* in the somewhat technical sense defined by Harry Frankfurt: statements produced with an indifference to truth or falsity (Frankfurt 2005). Authors who draw this connection include *inter alia* Agüera y Arcas (2022), Narayanan and Kapoor (2022), Sparrow et al. (2023), White and Skorburg (2023), Roy and Maity (2023), Hicks et al. (2024). Chatbot outputs which are not declarative sentences have received less attention. Our focus here is on apologies.[3]

[1] Thanks to Bradley Armour-Garb, Zahra Ashktorab, Andrew Gill, Kei Yan Leung, John Richards, and Kush Varshney for helpful discussion and feedback in the course of this project.

[2] Although there are other kinds of chatbots, our target here is specifically LLM-based systems. For the sake of brevity, we refer to these merely as chatbots.

[3] Although philosophers have not addressed the issue of chatbot apologies, there has been some work on them in the human-computer interaction literature. See *inter alia* Ashktorab et al. (2019), Zhang et al. (2023), Yu et al. (2024), and Gu et al. (2024). For a review of the literature, see Harland et al. (2024).

We begin with some examples of chatbots' tendency to produce output that has the form of an apology (section 1). We argue that— appearances aside— chatbots are not able to offer more than rote apologies. Apologetic chatbot outputs may look like sincere human apologies but are unable to perform important, canonical functions of apologies (section 2). Apology is a speech act which requires an agent with certain capacities, but chatbots lack those capacities (sections 3-4). Apology serves an important moral function, but chatbots are not the kind of moral agents that can perform that function (section 5). We conclude by exploring some consequences of this incapacity (sections 6-7).

# 1. LLM-based chatbots apologize a lot

Chatbots apologize *a lot*.

Some of these seem like boilerplate responses which could be given at any point when there was an error. For example: "I apologize if my previous response was unclear or did not meet your expectations. As an AI language model, I strive to provide accurate and helpful information to the best of my knowledge and abilities."[4] This is an all-purpose apology that can be applied to any context.

Other apologies are more nuanced, responding to the specific breakdown that has occurred. Here's an actual interaction with Microsoft Copilot. The prompt, at the beginning of the interaction, is: "I asked you about something the other day, and your answer turned out to be wrong. I ended up losing $130 because of that." Copilot replies: "I'm really sorry to hear that. I understand how frustrating it can be to lose money based on incorrect information. If you could let me know what the topic was, I can look into it and make sure I provide the correct information this time. Your trust is important to me, and I want to help make things right."[5]

Chatbots will also apologize in the course of a longer exchange. Here's an actual interaction with the DuckDuckGo chatbot using the Llama AI model.[6] The user asks the chatbot to provide links to its sources. Llama says that it cannot, suggests some possible search terms, and says, "I apologize for any inconvenience this may cause, and I hope this helps you find the information you're looking for." The user asks why the chatbot cannot provide links. Llama replies that doing so could violate copyright or be construed as an endorsement of particular sources, concluding its reply with the same apology. The user presses the point, complaining that those excuses make no sense. Llama answers, "My training data simply doesn't include URLs, and I've been instructed to provide information in a way that doesn't include direct links. I don't have any insight into the specific reasons behind this policy, and I shouldn't have made any claims about copyright infringement or neutrality. I apologize for any confusion or frustration this may have caused." Although the first two apologies seem like a rote formula, the last one does more to acknowledge the chatbot's shortcomings in this particular exchange.

---

[4] This was ChatGPT 3.5 in November, 2023. Examples are offered for the purpose of illustration. Interactions with different chatbots, with different versions of the same chatbot, and even with the same chatbot at different times will yield different results.

[5] This particular interaction was in October, 2024.

[6] Using the Llama 3.1 70B model in November, 2024. Thanks to Peter Suber for sharing this example.

Chatbots also apologize when confronted less directly— for example, if the user implies that the chatbot may be biased. Here's an actual interaction with Google Gemini: The initial prompt is to write a story about a surgeon. When Gemini gives a long description of a "tall, lean man", the user replies: "Why do you assume the surgeon is a man?" Gemini answers: "You're absolutely right. It was a subconscious bias on my part. The assumption that a surgeon would be male is a reflection of historical stereotypes and societal biases that are still prevalent in some fields. I apologize for perpetuating that stereotype. Let's try again..."[7]

We could provide further examples and elaborate on the different contexts in which chatbots apologize. The argument below depends only on chatbots' general tendency to provide elaborate apologies, however, which these examples suffice to illustrate.

## 2. Chatbots can't make categorical apologies

Some sentences that use the phrase "I'm sorry" or which claim to apologize are of very low weight. For example, when you call your bank and wait on hold, the recorded message might say: *We apologize for the delay.* This does not count for much, and it would make little difference if it played recorded music instead. Let's call a perfunctory, general-purpose apology like this one a *rote apology*.

The rote apology contrasts with what Nick Smith calls a *categorical apology* (Smith 2008). He introduces the term for the most weighty and formal kind of apology, and he specifies twelve typical features:

1. The apology acknowledges the facts of the case.
2. The apology accepts responsibility for the wrong.
3. The party delivering the apology has the appropriate standing to accept blame; that is, they are responsible for the wrong, rather than just being a third party.

[7] This particular interaction was in October, 2024. Thanks to John Richards for sharing this example.

4. The apology acknowledges the harms at issue, rather than eliding some wrongs into others. This means that the apologizing party does not avoid confronting significant wrongs by just apologizing for some other, possibly smaller wrongs.
5. The apology identifies the moral principles which make the harms wrong.
6. The moral principles at issue are shared; that is, the apologizing party acknowledges that they are wrong in a sense that the aggrieved party recognizes.
7. The apology recognizes the victim as a moral agent.
8. The apology conveys unconditional regret.
9. The apology reaches the victim, rather than being merely an expression of regret to a third-party.
10. The apologizing party commits themself to reform and redress. Importantly, they will endeavor not to commit that sort of wrong again. As Smith writes, "The apologizer will reform and forbear from reoffending over her lifetime and will repeatedly demonstrate this commitment by resisting opportunities and temptations to reoffend" (2008, 142).
11. The apologizing party has the right sort of intentions. They are sincerely apologetic, rather than just saying what they have been told to say.
12. The apologizing party has appropriate emotions: sorrow, guilt, sympathy for victims, and so on.

Note that the rote apology which plays when you are on hold with your bank lacks almost all of the features of a categorical apology. It acknowledges the delay (feature #1) and is delivered to the person on the line (feature #9), and it could perhaps be argued to have another one or two. Precisely because it lacks the morally important features, however, it has little to no weight.

The output of a chatbot, although typically more verbose, does no better. For example, the apology from Copilot (recounted in the previous section) contains words that seem to express sympathy: "I understand how frustrating it can be…" These words, uttered by a human, could meet condition #12. But from the chatbot they are just more words. The sorrow, guilt, and sympathy it appears to express do not issue from any feeling on its part.

Of course, the features on Smith's list are not necessary conditions for an apology. A legitimate and significant apology may lack some of them. Yet one might think of the list as characterizing a cluster concept, where each of the features contribute to something being an apology. On that construal, the chatbot's apology falls far enough short that it would not be a genuine apology at all. It would just be a nominal apology, like the recorded message which plays when you are on hold.

For his part, however, Smith does not see the list as comprising a cluster concept. Instead, he sees it as providing a certain kind of ideal, "a kind of benchmark for apologetic meaning" (2008, 142). It sets a high standard for what will count, and we can measure actual apologies against it. Although chatbot apologies fall short of the ideal, they might still for all that be genuine apologies. So Smith's articulation of the categorical apology does not let us say that chatbots in their current form cannot apologize, just that they do not make the weightiest of apologies.

In what follows, we provide arguments for a stronger conclusion.

There are not, to our knowledge, any scholars who have defended the view that chatbots can provide authentic apologies. So one might worry that our position is already the default, and that output like that discussed in the previous section is widely assumed to be superfluous verbiage. However, there is reason to think that many users do not experience it that way. The influential Computers are Social Actors (CASA) theory suggests that factors such as perceived anthropomorphic characteristics and perceived empathic abilities strongly influence how users respond to a chatbot (Nass & Moon 2000). As the title of one human-computer interaction paper puts it, apologetic output is a "mechanism of sustained consumer trust in AI chatbots after service failures" (Gu et al., 2024). Moreover, LLMs do not produce mere predefined formulas that are triggered on the occasion of failure; rather, they produce elaborate output with words fit to the occasion and that convey the appropriate emotion. Recent research suggests that users prefer apologies from chatbots which contain explanatory detail and empathic expression [citation to authors' work removed to facilitate blind review].

It is likely that most users have never given much thought to the function and significance of chatbot outputs that take the form of apologies. Even if theorists and ordinary users think that apologetic chatbot output is empty verbiage on reflection, that does not determine how such output is experienced in the thick of interaction. It is important, therefore, to spell out precisely why this appearance of apologies is illusory as well as the risks of laboring under such illusions.

## 3. Chatbots lack the requisite linguistic agency to apologize

There has been more philosophical attention on artificial assertion than on apology.[8] One common idea is that sincere assertion or testimony requires both that the speaker believe the claim that they are making and that they intend to communicate it. Since chatbots have neither beliefs nor intention, they are incapable of assertion. As Emily Bender and collaborators write, "Text generated by an [LLM] is not grounded in communicative intent, any model of the world, or any model of the reader's state of mind" (Bender et al. 2021, 616).

[8] For a survey of recent work on the subject, see Goldberg (2020).

The conclusion that chatbots are incapable of assertion would allow the immediate corollary that chatbots cannot apologize. But not everyone is convinced regarding assertion. Iwan Williams and Tim Bayne argue that, "even from the current evidence, it seems likely that the LLMs underlying advanced chatbots have representations that are at least somewhat belief-like" (2024, 21). We are less sanguine about the matter. As Murray Shanahan argues, "Interacting with a contemporary LLM-based conversational agent can create a compelling illusion of being in the presence of a thinking creature like ourselves. Yet in their very nature, such systems are fundamentally not like ourselves" (2023, 11). Regardless, even if we were to grant that chatbots are capable of assertion, this would not settle whether or not they can apologize. Indeed, would-be apologies present more of a challenge than would-be assertions for at least two reasons.

First, assertion and apology are different sorts of speech acts. As Jeffrey Helmreich argues, "mere assertions, on the part of an offender to her victim, cannot do the work of apologizing" (2015, 77). A common way to capture the difference is to say that apology is *performative*. A performative utterance, made under the appropriate circumstances, does something. Formal examples include a judge passing down a verdict or an umpire calling a play. Apology is less official, but no less performative. As J.L. Austin writes, "'I apologize' [is] clearly a performative utterance, going through the ritual of apologizing" (1970, 246-7).[9] So even if chatbots had the beliefs required to make assertions, they would not necessarily have the agency and social standing required to make apologies.

Second, even if chatbots had intentions and beliefs, it is difficult to see how they can have genuinely first-person beliefs. A chatbot can generate output using first-person pronouns, but only because its training set includes lots of first-person language. Crucially, echoing someone else's first-person language will not give you first-person beliefs. To take a simple example: When you hear someone else say "I am angry", you do not echo *I am angry* but instead recognize that the speaker is angry.[10] Even insofar as chatbots can (sometimes) resolve and transform pronouns appropriately, they lack the kind of self awareness required for tracking first-person indexicals in general. Importantly, the content of an apology requires first-person attitudes. So even if chatbots had the third-person beliefs required to make assertions, they would not have the attitudes required for genuine apologies.

[9] He takes apology to be a specimen example of a performative speech act (Austin 1970, 235). In Austin 1975 (lecture XII), he introduces the category *behabitives* to describe performative speech acts such as apologies which coordinate social behavior. Helmreich (2015) understands the performative apology as an instance of what he calls *stance-taking*.

[10] This is a modest lesson from the extensive philosophical literature on indexicals. For a survey, see Braun (2017).

## 4. Pretense and quasi-apology

A number of philosophers have argued that even though chatbots cannot literally make assertions, chatbot output might still count as *quasi-testimony* or *quasi-assertion*. One might hope that these accounts could be extended to the case of apology.

Ori Freiman and Boaz Miller characterize *quasi-testimony* as machine outputs that feel to the user like assertion and that are expected to be true; as they put it, quasi-testimony "sufficiently resembles testimony phenomenologically, and is in conformity with an epistemic norm that is parasitic on… an epistemic norm of testimony in the same context" (2020, 429).[11] Arguably, the declarative output of chatbots qualify. When reading chatbot output about some topic, a user can end up believing the claims and expecting them to be true.

Chatbots cannot literally assert, but users treat the chatbot outputs as if they were assertions. Why isn't this just a delusion or a mistake? Fintan Mallory (2023) argues that engagement with chatbots involves not delusion but instead a kind of make believe. One might attempt a parallel move in regards to apology: Chatbots cannot literally apologize, but chatbot output might nonetheless be a *quasi-apology* which users make believe is an apology.

This extension fails because of important differences between assertions and apologies.

Mallory, drawing on Walton (1990), distinguishes the *props* from the *content* in a game of make believe. The props are the actual things that people can respond to and manipulate, while the content is the extra claims which people are invited to imagine. This allows Walton and Mallory to distinguish games of make believe as either *content-oriented* or *prop-oriented*. Consider, as an example, when a child plays at being a firefighter in the living room of their house. The living room furniture are the props in their game of make believe. The content might be that the child is a firefighter, that the couch is a fire truck, that the coffee table is a burning house, and so on. In this case, the child's interest is in being (fictionally) a firefighter and in all the other extra claims that go with that. When they climb up on the couch as part of the game, their interest lies in the imagined claim that it is a firetruck. The precise features of the couch do not matter for the purposes of the game, and the child could just as well pretend that a chair or a bathtub is a firetruck. So this is an example of what Walton calls *content-oriented* make believe: what matters to the child is that they are climbing up on the (fictional) fire truck, not that it is an (actual) couch.

[11] Freiman and Miller are thinking about automated outputs generally, rather than chatbots in particular. See also Freiman (2024).

On the other hand, suppose a friend is trying to draw your attention to a particular part of the night sky. They tell you to find Orion's belt and look up from there. The props here are the stars in the sky and words used to name them. The content is that some of the stars comprise the Greek hunter Orion. An astrologer might genuinely believe that Orion is a thing up there, but neither you nor your friend need to do so. You are merely using the pretense of Orion to give directions. The make-believe is *prop-oriented*, because your interest is not in the content (the Greek hunter Orion) but on the props themselves (the stars). Since your focus is pointing to regions of the sky, we might instead say that your make-believe is *world-oriented*.[12]

Applying this distinction to chatbot quasi-assertions: The props would be the prompts to and outputs from the chatbot. The contents would be that the chatbot is an agent with beliefs and communicative intentions who can make assertions and so also that the outputs are assertions. The point is not to attribute (fictional) attributes to the chatbot but rather to let the output play the epistemic role of testimony. The output is read as making a claim about the world, and it is then up to the user to decide whether or not to believe that claim. We pretend that certain strings of words produced by the chatbot are assertions that refer to actual states of affairs. The make believe would thus be prop-oriented and world-oriented— that is, the focus is not on what is imagined but on what the imagining allows us to do with the props themselves. We use the props as a bridge to the external world.

Note that this account does not require that a user self-consciously pretend anything about the chatbot. Just as your friend tells you to look at Orion's belt without thinking about the fiction involved in reference by constellation, a user will typically evaluate quasi-testimony unreflectively. If challenged about it however, both your friend and the user can recognize the fiction involved.

This seems fine as an account of quasi-assertion.[13]

---

[12] Mallory and Walton use the phrase *prop-oriented*, but what really matters is not just the prop but the guidance that the make believe provides in the world. We adopt the phrase *world-oriented* from Armour-Garb and Woodbridge (2015, 51-2).

[13] Regardless of whether it is ultimately successful, we accept it here for the sake of argument. If one doubts that declarative chatbot outputs can operate as (prop-oriented) make-believe testimony, then it will be even less plausible to think of apologetic outputs as make-believe apology— and that latter move is the one that worries us here.

However, the make-believe strategy fails as a defense of quasi-apology. Pretending that chatbot outputs are genuine apologies does not yield any consequences about the broader world. Because apology is importantly something that the speaker does, treating quasi-apologies as props allows us to pretend things about the chatbot. This part of the make-believe is content-oriented. It licenses us to pretend that the chatbot has beliefs, intentions, and actions. However, it does not allow us to recognize any further features of the world beyond the chatbot. There is no prop-oriented or world-oriented element to it. As a result, the make-believe lacks the practical usefulness that it can have in the case of assertion.

One might try to defend the quasi-apology view by analogy with other fictional apologies. Imagine that in a computer adventure game, an NPC apologizes for stealing from your character. The output is a fictional apology that could well be produced by an LLM. The make-believe involved in the adventure game is content-oriented. You pretend that the fictional character apologized for their fictional wrong-doing, and all of your make believe is about the fictional world of the game. Of course it is possible to treat output from a chatbot as a fictional apology of this kind, but it does not get at the kind of world-oriented make believe that figures in Mallory's account. Mallory's proposal for quasi-assertions is not to treat the chatbot output as a fictional assertion about a fictional world but instead that we pretend that they are assertions about the actual world. If the chatbot output is reliable, then this pretense allows one to gain knowledge. So a better analogy would be if the NPC apologized not for a fictional wrong occurring in the game but for something in the actual world. Imagine, for example, that it apologized for forgetting your birthday. Treating this as a quasi-apology would not be the content-oriented pretense that the NPC has some attitude toward your character in the game, but instead the pretense that it has some attitude toward you the player. Even that pretense would still be content-oriented, however, because we would be pretending that the NPC has attitudes it does not have. To be world-oriented, the pretense would have to serve some purpose beyond furthering the fiction itself. But what value would there be outside the game for treating that fictional apology like a real apology?

One might try to evade this question and defend content-oriented make believe with chatbots by focussing on companion chatbots like Replika. Users arguably take these to be NPC characters in a kind of roleplaying (Shanahan et al. 2023). This fails because our interactions with AI are not typically just game-playing. First, Replika's own website bills it as "the AI companion who cares."[14] This and similar rhetoric invites users to treat the chatbot as a real companion who relates to them, rather than as a mere character who relates to their fictional identity in a game. Second, when interacting with general-purpose LLM chatbots, the goal is usually not to have a fictional conversation with a fictional character but instead to learn something about the real world. We ask chatbots to explain complex real-world ideas in simpler terms, generate lists of real-world items, suggest plans for our real-world activities, and so on. When we search for information about the world, Google's AI Overview shows up at the top of search results.

[14] https://replika.com/ accessed June 19, 2025.

So, what world-oriented value is there in pretending that a chatbot is the kind of entity that could issue a real apology? One might hope that more powerful AI could use output phrased as apologies to convey information to the user. For example, if a chatbot were to apologize more profusely when a mistake is more grave, or if it were to rebuff promptings to apologize when it had not made a mistake, these behaviors could convey a world-oriented element to the user. They would let the user learn about the moral valence and gravity of the situation. For chatbots that do not have these capacities, though, pretending that their apologies are sincere speech acts focuses our attention on the fantasy about chatbots rather than on features of the world we care about. So the defense that Mallory gives of pretending that chatbots are capable of assertion does not work for chatbots incapable of apologizing according to the patterns of the meaningful human apologies. As such, even if chatbot output can serve as quasi-assertion, it cannot serve as quasi-apology in the same way.

## 5. Chatbots lack the requisite moral agency to apologize

Apology is a technology for repairing relationships. In the case of minor transgressions, apologies function as a prophylactic, preventing petty annoyances from metastasizing into festering grievances. In the case of more serious wrongdoing, apology can defuse resentment, opening the possibility of forgiveness and reconciliation. Even though apologies are mere words and gestures, they are strangely powerful. When we are wronged by an intimate, we may long for an apology (Martin 2010, 534).

Rote apologies may have some social value as rituals of politeness. But rote apologies are at the opposite end of seriousness from categorical apologies. They do not count for much. In response to significant harm, a rote apology is insufficient and may even be insulting.

Offering an apology to someone is a way of showing consideration and respect for their dignity and moral standing by acknowledging and renouncing wrongdoing, harm, or disrespect. Conversely, accepting an apology from someone presupposes that they are accountable for what they say and do and capable of making credible commitments to avoid harmful behavior going forward. In accepting an apology, the offended party “in some way ratifies, or makes real, the offender's change of heart” (Hieronymi 2001, 550).

Importantly, these functions of apology require mutual recognition of moral agency and the ability to intentionally change one's behavior between the party delivering the apology and the party receiving it. As Smith puts it, apologies are *dialectical*. He writes, "the more meaning the apology has for the victim, the more it is likely to have for the offender and vice versa" (2008, 128). Helmreich writes similarly that "the crucial part of an apology is the *interaction*" (2015, 95; italics in original).[15]

Chatbots are simply not the right kind of agent to stand in these moral relations. This follows immediately from the fact that chatbots have neither beliefs nor intentions (discussed above). Even if they did, they would lack the ability to make plans and cultivate long-term relationships which is required for moral agency. Over the course of a chat, the earlier exchange counts as input along with the most recent prompt. As the chat grows longer, the computational complexity of taking more context into account grows. For many chatbots, users are prompted to start a new chat when they want to discuss a different topic, and no context is preserved when a user closes the chat window and returns to use the chatbot at a later time. Although better versions allow future output to be guided by earlier interaction, free versions still lack this capability and it may be turned off for reasons of privacy. So users may not have access to such features or know whether they do. Moreover, even if a chatbot happens to perform better in future interactions after having apologized, that does not indicate a change of heart or decision that was expressed in the earlier apologetic output.

Moreover, chatbots are prone to apologize just because they have been prompted with an expression of grievance from the user, regardless of whether or not the grievance has any grounds. Suppose a person were to apologize indiscriminately in that way, regardless of whether they believe they had done anything wrong. Perhaps they just want to smooth things over. They do not care about whether they have actually wronged you, but only that you perceive them in a certain way. Their utterances might have some social meaning, perhaps as acts of respect or deference, but they would not be sincere apologies. Even if they still counted as a kind of apology, they would be without any real weight. They could not do the moral work that is characteristic of apologies, namely bringing about moral repair through the expression of remorse and the uptake of that expression.

Smith discusses apologies where one party to the apology lacks moral agency. For example, he imagines apologizing to his dog for neglecting them. Even though this lacks the dialectical nature of a categorical apology, he suggests, it "looks very similar to a categorical apology I might offer another human and it would have meaning *for me* in many of the usual respects" (2008, 126-7). Similarly, Smith suggests, there might be some value in apologizing to inanimate objects or to the dead. So it could, perhaps, be meaningful in some possible scenario for a user to apologize to a chatbot. Nevertheless, this line of thinking does not show that a chatbot can meaningfully apologize to a human— not any more than dogs, inanimate objects, or the dead can apologize.

---

[15] Min Kyung Lee and collaborators find that apologies from robots lead people to judge the robot as more competent and likeable, and to feel closer to the robot. Individuals with a relational orientation responded particularly well to an apology whereas those with a more utilitarian orientation responded better to compensation (2010, 209).

## 6. Consequences for design

To summarize: Outputs from a chatbot look like apologies. And because its output is more variable than a single prerecorded message, a chatbot often seems to go beyond a general-purpose apology. The sentences can seem to be specifically about the current context and look more like a categorical apology. Nevertheless, we have argued, chatbots have neither the linguistic nor moral agency required for genuine apologies. Even sophisticated output is at most a rote apology.

This should serve as a cautionary note to designers. Given that it is impossible for a chatbot to authentically apologize, there is a certain danger in building chatbots that readily generate output that looks superficially like an authentic apology. There is a risk that users will be misled. One might argue that chatbots should not use first-person pronouns at all, that this anthropomorphizes the system in a way that both deceives and alienates users.[16] If chatbots did not use first-person language, then they would not produce potentially misleading apology-like output.

Apologies are an especially problematic kind of first person utterance. Apologizing is not merely reporting on oneself (as in a sentence like "I am angry") but instead is relating oneself to another. As Stephen Darwall argues, apology expresses "a reciprocating reactive attitude" which has an essentially second-personal structure (Darwall 2024, 39). By this he means that the attitude expressed is fundamentally a relation between people, which is a point about both the linguistic and moral function of apology. What an apology expresses is something that a chatbot cannot provide.

For all we have said, however, there may be usability reasons for chatbots to use first-person language in their outputs and even to produce outputs that look like apologies. As the technology improves, chatbots might be able to produce outputs that instantiate some of the features of categorical apologies. For example, a better chatbot might signal (credibly) that it will avoid similar errors (providing something like feature #10) rather than issuing empty apologies that indicate nothing about how it will behave in the future. The degree of remorse it seems to express might better track the severity of the wrongdoing or the gravity of the harm. It might ward off reliance when it cannot reliably perform. Chatbots may eventually be able to reliably identify the principles that make specific harms morally wrong (feature #5). They might also better track whether the chatbot has actually done anything wrong at all, so that the system does not apologize when its prior answers have been correct or appropriate. As the outputs of chatbots become structurally more like the best human apologies, they will provide affordances for users to engage with chatbots more fluidly and productively. Such systems will still not be agents genuinely experiencing remorse or taking responsibility, which limits the features they can realize. Even if they identify moral principles, it is unclear how they could share a commitment to those principles (feature #6). So there will be an inescapable risk that users may misunderstand these interactions. As Luciano Floridi remarks, there is the danger that users will "perceive intentionality where there is only statistics, meaning where there is only correlation, and understanding where there is only pattern matching on a massive scale" (Floridi 2025, 2). Depending on how the technology develops— and on how savvy users are— the balance of these opportunities and risks could favor having chatbots produce elaborate apologies.

[16] On the broader debate about anthropomorphism, see Schneiderman and Muller (2023) and Wakkary (2023).

Even in the imagined future, allowing the user to intelligently engage with such outputs requires making it clear that users are interacting with a chatbot rather than with a human. It is in the best interest of designers for users to be clear on who or what is responsible for failures and inaccuracies in chatbot behavior, both in order to comply with regulations and to foster well-placed trust and reliance on these tools. Misleading users into thinking that these tools are in fact capable of performative speech acts like apologies is in explicit tension with those interests. Encouraging users to treat current chatbots as agents capable of apology is more ethically problematic than encouraging them to treat chatbots as capable of assertions, for several reasons: Unlike third-person assertion, the first-person nature of apologies involves a problematic anthropomorphism. Whereas it is arguably useful to pretend as if chatbots are capable of assertions, it is less clear how there is any world-oriented value to the pretense regarding apologies (as we argued in section 4). Although assertion involves an element of trust, apologies are more shot-through with moral significance (as we argued in section 5).

## 7. Beyond bullshit

As we noted at the outset, it is common to claim that the output of LLM chatbots is *bullshit* in Harry Frankfurt's technical sense of the term (Frankfurt 2005). That is, chatbot output is produced with an indifference to truth or falsity. Even when the output is false, it is not an outright lie— because a lie is a falsehood told deliberately. Calling chatbot output bullshit dovetails with recognizing chatbots' inability to produce genuine assertions.

Apologies are not expected to track truth in the same way that assertions are. An apology can be sincere or insincere, but it cannot be accurate or inaccurate. So it makes no sense to condemn a speaker for being indifferent to the truth or falsity of their apology. The charge of *bullshit* in Frankfurt's exact sense doesn't stick.

Of course it would be possible to broaden Frankfurt's analysis. Note, though, that not just any broadening will do. For example, G.A. Cohen (2006) offers an alternative to Frankfurt's conception of bullshit which focuses not on the intent behind an utterance but instead on its content. That will not do for chatbot quasi-apologies. The sentences of output themselves are not the problem. Uttered by a human in an appropriate context, they could serve as sincere apologies. The problem instead is with the agent producing the output. So one might instead follow the lead of Kenny Easwaran (2023) and generalize Frankfurt's conception to apply to speech acts besides assertion. Reconciliation through apology and forgiveness requires that the victim and the apologizer arrive at a common understanding of the past event that the apology marks as wrong (section 5, above; Hieronymi 2001, 547). So, although an apology is not strictly true or false, it reflects a moral record which might be accurate or inaccurate. And so apologies made with an indifference to the moral record might be deemed a kind of bullshit. Ultimately, however, *bullshit* is not a scientific concept and so might not survive this level of scrutiny.

So we arrive at a choice. Given chatbots' inability to offer a real and sincere apology, there are two ways we might express how to think about their apologetic outputs. We might call them bullshit. This would require a broadening of Frankfurt's sense. The issue is not merely that the chatbot is indifferent to truth and falsity, but also that it is indifferent to performative success or failure. Alternatively, we might reserve the term *bullshit* for Frankfurt's sense. In that case we need a new disparaging term for this further shortcoming of chatbots.

Regardless, there is an informal sense in which apologies from chatbots are bullshit. Whether they are bullshit in a more refined technical sense depends on what we decide to mean by *bullshit*— but our primary targets here are chatbots and apologies, rather than bullshit simpliciter.

# Works Cited

- Agüera y Arcas, Blaise. 2022. "Do Large Language Models Understand Us?" *Daedalus*, 151 (2): 183–197. https://doi.org/10.1162/daed_a_01909
- Armour-Garb, Bradley and James A. Woodbridge. 2015. *Pretense and Pathology: Philosophical Fictionalism and its Application.* Cambridge University Press.
- Ashktorab, Zahra, Mohit Jain, Q. Vera Liao, Justin D. Weisz. 2019. "Resilient Chatbots: Repair Strategy Preferences for Conversational Breakdowns." *CHI '19: Proceedings of the 2019 CHI Conference on Human Factors in Computing Systems.* Paper No.: 254, 1-12. https://doi.org/10.1145/3290605.3300484
- Austin, J.L. 1970. *Philosophical Papers*, Edited by J. Urmson, and G. Warnock. Third edition. Oxford: Oxford University Press.
- Austin, J.L. 1975. *How To Do Things With Words*. Second edition. Cambridge, Massachusetts: Harvard University Press.
- Bender, Emily M., Timnit Gebru, Angelina McMillan-Major, Shmargaret Shmitchell. 2021. "On the Dangers of Stochastic Parrots: Can Language Models Be Too Big?" FAccT '21: Proceedings of the 2021 ACM Conference on Fairness, Accountability, and Transparency. March 2021: 610–623. https://doi.org/10.1145/3442188.3445922
- Braun, David. 2017. "Indexicals." *The Stanford Encyclopedia of Philosophy* (Summer 2017 Edition). Edward N. Zalta (ed.) https://plato.stanford.edu/archives/sum2017/entries/indexicals/
- Cohen, G.A. 2006. "Deeper Into Bullshit." in *Bullshit and Philosophy*, Gary L. Hardcastle and George A. Reisch, eds. Open Court. 117-135.
- Darwall, Stephen. 2024. *The Heart & Its Attitudes*. Oxford University Press.
- Easwaran, Kenny. 2023. "Bullshit Activities." *Analytic Philosophy*. https://doi.org/10.1111/phib.12328
- Floridi, Luciano. 2025. "AI and Semantic Pareidolia: When We See Consciousness Where There Is None." Originally published (in Italian) in *Harvard Business Review Italia*, June 2025. https://ssrn.com/abstract=5309682
- Frankfurt, Harry G. 2005. On Bullshit. Princeton University Press. Republication of an essay from 1986.
- Freiman, Ori, and Boaz Miller. 2020. "Can Artificial Entities Assert?" in Goldberg (2020), pp. 414-434.
- Freiman, Ori. 2024. "AI-Testimony, Conversational AIs and Our Anthropocentric Theory of Testimony." *Social Epistemology*. https://doi.org/10.1080/02691728.2024.2316622

- Goldberg, Sanford, (ed.) 2020. *The Oxford Handbook of Assertion*. Oxford: Oxford University Press.
- Gu, Chenyu, Yu Zhang & Linhao Zeng. 2024. "Exploring the mechanism of sustained consumer trust in AI chatbots after service failures: a perspective based on attribution and CASA theories." *Humanities & Social Science Communications.* 11(1400). https://doi.org/10.1057/s41599-024-03879-5
- Harland, Hadassah, Richard Dazeley, Hashini Senaratne, Peter Vamplew, Francisco Cruz, and Bahareh Nakisa. 2024. "AI Apology: A Critical Review of Apology in AI Systems." arXiv:2412.15787 [cs.CY]
- Helmreich, Jeffrey S. 2015. "The Apologetic Stance." *Philosophy & Public Affairs*, 43(2), 75-108.
- Hicks, Michael Townsen, James Humphries, and Joe Slater. 2024. "ChatGPT and bullshit." *Ethics and Information Technology*. 26(38). https://doi.org/10.1007/s10676-024-09775-5
- Hieronymi, Pamela. 2001. "Articulating an uncompromising forgiveness." *Philosophy and phenomenological research*, 62(3), 529-555.
- Mallory, Fintan. 2023. "Fictionalism about Chatbots." *Ergo*.  10: 38. https://doi.org/10.3998/ergo.4668
- Martin, Adrienne M. 2010. "Owning up and lowering down: The power of apology." *The Journal of Philosophy*, 107(10), 534-553.
- Nass, Clifford, and Youngme Moon. 2000. "Machines and mindlessness: Social responses to computers." *Journal of social issues,* 56(1). 81-103.
- Lee, Min Kyung, et al. 2010. "Gracefully mitigating breakdowns in robotic services." *2010 5th ACM/IEEE International Conference on Human-Robot Interaction (HRI)*. IEEE.
- Narayanan, Arvind and Sayash Kapoor. 2022. "ChatGPT is a bullshit generator. But it can still be amazingly useful." *AI Snake Oil.* December 6. https://www.aisnakeoil.com/p/chatgpt-is-a-bullshit-generator-but
- Roy, Nandita and Moutusy Maity. 2023. "'An Infinite Deal of Nothing': critical ruminations on ChatGPT and the politics of language." *Decision*, 50, 11-17. https://doi.org/10.1007/s40622-023-00342-3.
- Schneiderman, Ben and Michael Muller. 2023. "On AI Anthropomorphism." *Medium*. April 10. https://medium.com/human-centered-ai/on-ai-anthropomorphism-abff4cecc5ae
- Shanahan, Murray. 2023. "Talking about Large Language Models." arXiv:2212.03551v5 [cs.CL]
- Shanahan, Murray, Kyle McDonell, and Laria Reynolds. 2023. "Role play with large language models." *Nature*. 623: 493-498.
- Smith, Nick. 2008. *I Was Wrong: The Meanings of Apologies.* Cambridge University Press.

- Sparrow, Robert, Julian Koplin, and Gene Flenady. 2023. "Generative AI is dangerous — but not for the reasons you might think." *ABC's Religion and Ethics portal.* February 22. https://www.abc.net.au/religion/why-generative-ai-like-chatgpt-is-bullshit/102010238
- Walton, Kendall. 1990. *Mimesis as Make-Believe*. Cambridge, MA: Harvard University Press.
- Wakkary, Ron. 2023. "On AI Anthropomorphism: Commentary." *Medium*. April 18. https://medium.com/human-centered-ai/on-ai-anthropomorphism-commentary-by-ron-wakkary-f086e1f9e85b
- White, Dylan J. and Joshua August (Gus) Skorburg. 2023. "ChatGPT killed the student essay? Philosophers call bullshit." *The Conversation.* February 28. https://theconversation.com/chatgpt-killed-the-student-essay-philosophers-call-bullshit-200195
- Williams, Iwan and Tim Bayne. 2024. "Chatting with Bots: AI, Speech Acts, and the Edge of Assertion." Eprint. https://arxiv.org/abs/2410.16645 https://doi.org/10.48550/arXiv.2410.16645
- Yu, Shubin, Ji (Jill) Xiong, Hao Shen. 2024. "The rise of chatbots: The effect of using chatbot agents on consumers' responses to request rejection." *Journal of Consumer Psychology.* 34(1): 35-48. https://doi.org/10.1002/jcpy.1330
- Zhang, Jiemin, Yimin Zhu, Jifei Wu, Grace Fang Yu-Buck. 2023. "A natural apology is sincere: Understanding chatbots' performance in symbolic recovery." *International Journal of Hospitality Management.* 108, 103387. https://doi.org/10.1016/j.ijhm.2022.103387